\documentclass[preprint2]{aastex631}

\shorttitle{Transition region beneath active region upflows}
\shortauthors{Huang et al.}
\def \kms {{\rm $km\;s^{-1}$}}

\def \arcsec {$^{''}$}
\def \siiv {Si\,{\sc iv}}
\def \cii {C\,{\sc ii}}

\def \mgii {Mg\,{\sc ii}}

\newcommand{\halpha}{H$\alpha$}
\graphicspath{{./}{figures/}}
\begin{document}

\title{Dynamics in the transition region beneath active region upflows viewed by IRIS}

\correspondingauthor{Zhenghua Huang}
\email{z.huang@sdu.edu.cn}

\author{Zhenghua Huang}
\affiliation{Shandong Key Laboratory of Optical Astronomy and Solar-Terrestrial Environment,\\
Institute of Space Sciences, Shandong University, 264209, Weihai, Shandong, China}

\author{Lidong Xia}
\affiliation{Shandong Key Laboratory of Optical Astronomy and Solar-Terrestrial Environment,\\
Institute of Space Sciences, Shandong University, 264209, Weihai, Shandong, China}

\author{Hui Fu}
\affiliation{Shandong Key Laboratory of Optical Astronomy and Solar-Terrestrial Environment,\\
Institute of Space Sciences, Shandong University, 264209, Weihai, Shandong, China}

\author{Zhenyong Hou}
\affiliation{School of Earth and Space Sciences, Peking University, Beijing 100871, China}

\author{Ziyuan Wang}
\affiliation{Shandong Key Laboratory of Optical Astronomy and Solar-Terrestrial Environment,\\
Institute of Space Sciences, Shandong University, 264209, Weihai, Shandong, China}



\begin{abstract}
Coronal upflows at the edges of active regions (AR), which are a possible source of slow solar wind, have been found to connect with dynamics in the transition region. 
To infer at what scale transition region dynamics connect to AR upflows, we investigate the statistical properties of the small-scale dynamics in the transition region underneath the upflows at the edge of AR NOAA 11934.
With observations from the Interface Region Imaging Spectragraph (IRIS),
we found that the \siiv\,1403\,\AA\ Doppler map consists of numerous blue-shifted and red-shifted patches mostly with sizes less than 1\,$Mm^2$.
The blue-shifted structures in the transition region tend to be brighter than the red-shifted ones, but their nonthermal velocities have no significant difference.
With the SWAMIS feature tracking procedure, in IRIS slit-jaw 1400\,\AA\ images we found that dynamic bright dots with an average size of about 0.3\,$Mm^2$ and lifetimes mostly less than 200\,s spread all over the region.
Most of the bright dots appear to be localised, without clear signature of propagation of plasma to a long distance on the projection plane.
Surge-like motions with speeds about 15\,\kms\ could be seen in some events at the boundaries of the upflow region, where the magnetic field appear to be inclined.
We conclude that the transition region dynamics connecting to coronal upflows should occur in very fine scale, suggesting that the corresponding coronal upflows should also be highly-structured.
It is also plausible that the transition region dynamics might just act as stimulation at the coronal base that then drives the upflows in the corona.
\end{abstract}


\section{Introduction} \label{sec:intro}
The edge of solar active region, where the magnetic topology might be open\,\citep{2014A&ARv..22...78W}, has been recognized as a candidate of source regions of the slow solar wind\,\citep[e.g.][etc.]{2007Sci...318.1585S,2011ApJ...727L..13B,2012SoPh..281..237V,2013SoPh..283..341D,2013SoPh..286..157S,2014SoPh..289.3799C,2015NatCo...6.5947B,2015SoPh..290.1399F,2015ScChD..58..830Z,2017PASJ...69...47H}.
These regions are usually identified as plasma upflows on the Doppler velocity maps measured with coronal spectral lines\,\citep[e.g.][]{2004A&A...428..629M,2008ApJ...676L.147H,2008ApJ...686.1362D,2015NatCo...6.5947B}.
The line-of-sight velocities of these plasma upflows in the corona are normally in the range of 10--50\,\kms\,\citep[for summaries, see reviews by][]{2016SSRv..201...55A,2019PASJ...71R...1H,2021SoPh..296...47T}. 

\par
%Where active region upflows are initiated has been investigated intensively.
Many studies\,\citep[e.g.][]{2011ApJ...727L..13B,2012ApJ...760L...5B,2014SoPh..289.3799C,2015NatCo...6.5947B} found that the relative abundance of low first ionisation potential (FIP) elements of the active region upflows is enhanced significantly, indicating a source region different from the photosphere\,\citep{2017ApJ...836..169F}.
\citet{2004A&A...428..629M} found a consistency between upflow features in C\,{\sc iv} (with a formation temperature of 0.1\,MK) and those in Ne\,{\sc viii} (with a formation temperature of 0.6\,MK), which suggests the upflows might be started in a place where C\,{\sc iv} is formed (i.e. mid-transition-region) or lower.
The existence of upflows in the transition region of open field regions have also been confirmed by observations of O\,{\sc iv} lines\,\citep{2008ApJ...685.1262M}.
Using both imaging and spectral data, \citet{2010A&A...516A..14H} found in a case that the plasma flows upward along a strand intermittently with chromospheric jets occurring at its root, suggesting a chromospheric origin of the upflows.
Similarly, \citet{2010ApJ...722.1013D} suggest that chromospheric jets that are rapidly heated to coronal temperatures at low heights are associated with quasi-periodic upflows along coronal loops\,\citep{2010RAA....10.1307G,2011ApJ...727L..37T,2012ApJ...759..144T,2016ApJ...825...58R}.
A study by \citet{2011ApJ...737L..43N} also shows a similarity between the spectral profiles of a jet and those of the upflow, and thus it suggests that the upflow might originate in the lower corona via explosive processes.
A case study on interaction between an EIT wave and a region of active region upflows suggests that the upflows can be disturbed by activities in various layer of the solar atmosphere\,\citep{2011ApJ...740..116C}.

\par
A further study involving \halpha\ images suggests that chromospheric jets alone cannot supply enough materials to the active region upflows\,\citep{2015A&A...584A..38V}.
The phenomenon might be resulted from both magnetic-reconnection-induced jets and pressure-induced flows in the lower solar atmosphere\,\citep{2014Ap&SS.351..417L,2014SoPh..289.4501S,2015A&A...584A..39G} and it requires even more complicate processes to allow the plasma to escape to the slow solar wind\,\citep{2013SoPh..283..341D,2014SoPh..289.4151M,2017SoPh..292...46B}.
On another hand, \citet{2011ApJ...727...58W} found that the upflow region can have complex velocity structures, including the outflow regions with little or no emission in Si\,{\sc vii} and those along fan loops with downflows in their cooler footpoints\,\citep[see also][]{2011ApJ...730...37U,2012ApJ...744...14Y}.
Such complex velocity structure in active region upflows has also been confirmed from the aspects of images\, \citep{2012ApJ...752...13B}, density distributions\,\citep{2015ApJ...805...97K} and field extrapolations\,\citep{2016SoPh..291..117E}.

\par
Recently, by analysing coordinated data from Hinode and High-resolution Coronal Imager (Hi-C), \citet{2020ApJ...894..144B} found that the emission of the active region upflow region contain two components, one from expand plasma related to expelled close loops and the other one from the dynamic activity in the plage region.
More importantly, the abundance of the emission component from the dynamic activity in the plage region is similar to that of the photosphere\,\citep{2020ApJ...894..144B}.
The lower solar atmosphere origin has also been confirmed by a very recent study on the upflows in a newborn active region\,\citep{2021arXiv210603318B}.

\par
Based on high resolution spectral observations from the Interface Region Imaging Spectrograph (IRIS), \citet{2020ApJ...903...68P} analysed in detail the spectral data of Mg\,{\sc ii} (low chromosphere line), C\,{\sc ii} (mid chromosphere line) and \siiv\ (low transition region line) underneath two upflow regions.
Their statistics indicates that the probabilities of blue-shifted pixels in the \cii\ and \siiv\ observations of upflow regions are significantly higher than those from a nearby moss regions.
They also found signature of chromospheric upflows in the asymmetries of \mgii\ lines emitted from the upflow regions.
They concluded that the atmosphere (from low chromosphere to corona) of the upflow regions should be treated as an interconnected system.

\par
Here, we analyse an IRIS data set of active region upflows, in which we observe clear velocity and intensity structures in the transition region.
Thanks to the unprecedentedly-high resolution of the data, we can measure the properties of the dynamic structures in the transition region.
This will allow us to infer quantitatively at what scale the transition region dynamics are connecting to the coronal upflows,
and thus help understand how the activities in the lower atmosphere link to the upflows in the corona.

%%%%
\begin{figure*}[ht!]
\includegraphics[width=\textwidth,clip,trim=0.2cm 0cm 0cm 0cm]{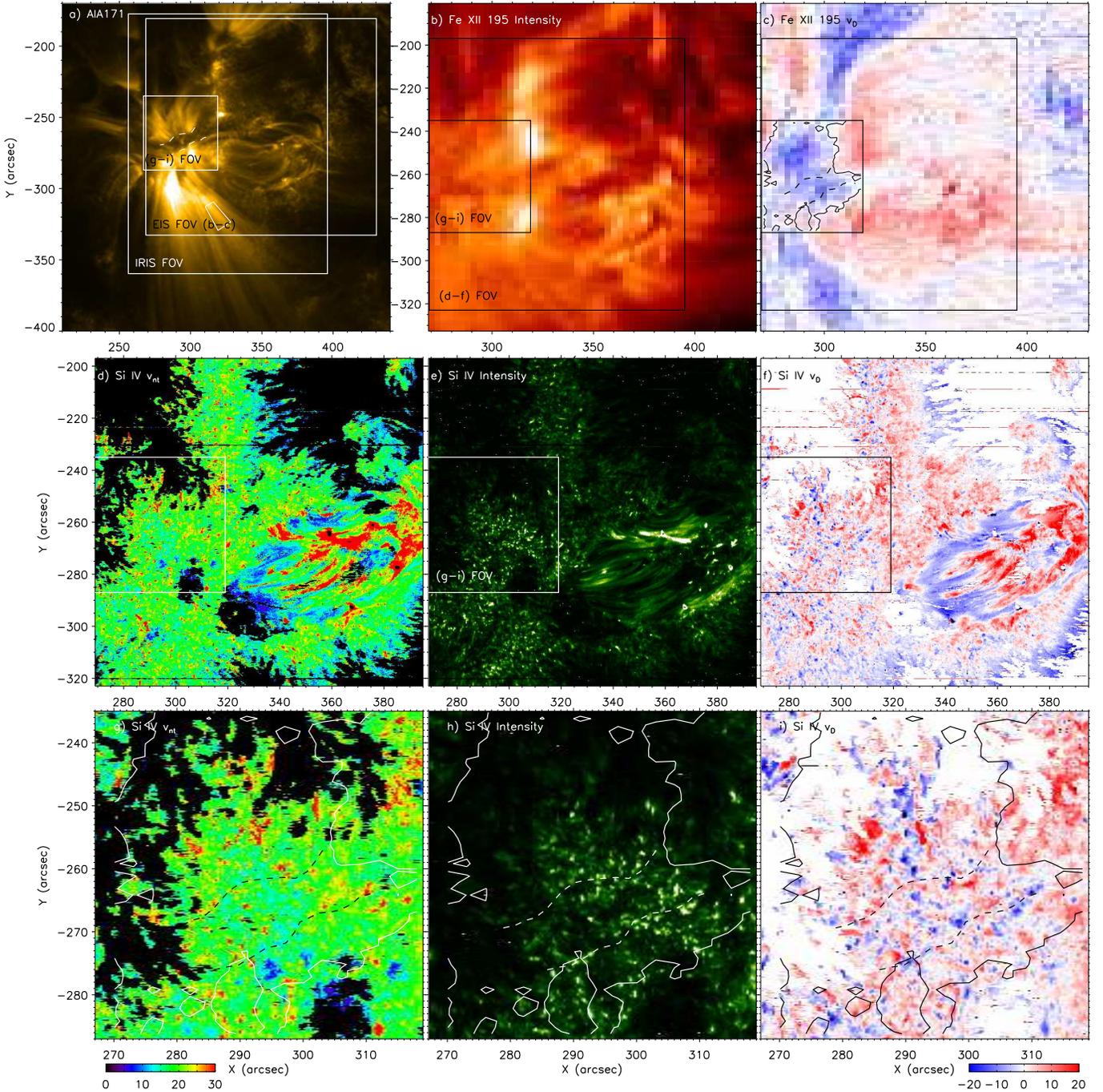}
\caption{The region of interest seen in AIA\,171\,\AA\ (a), EIS Fe\,{\sc xii}\,195\,\AA\ intensity (b) and Doppler maps (c), IRIS Si\,{\sc iv}\,1403\,\AA\ nonthermal velocity (d), intensity (e) and Doppler maps (f).
The EIS and IRIS FOVs are indicated in panel a.
The active region upflows can be identified as a blue-shifted region on the left part of the EIS Fe\,{\sc xii} Doppler map (panel c).
The core of the upflow region analysed in detail is denoted in panels a--f, and its IRIS Si\,{\sc iv}\,1403\,\AA\ nonthermal velocity, intensity and Doppler velocity images are zoomed-in and shown in panels g--i.
The contours of Fe\,{\sc xii} Doppler velocity at zero are shown in panels c and g--i.
The dashed lines in panels (a), (c), (g)--(i) denote the locations of bright feet of fan loops as seen from the AIA\,171\,\AA\ image.
The units of nonthermal velocities and Doppler velocities are \kms.
The white trapezoid in panel (a) denotes a typical fan loop in the active region.
\label{fig:fov}}
\end{figure*}
%%%%

\begin{figure*}[ht!]
\includegraphics[clip, trim=0cm 8cm 0cm 0cm,width=\textwidth]{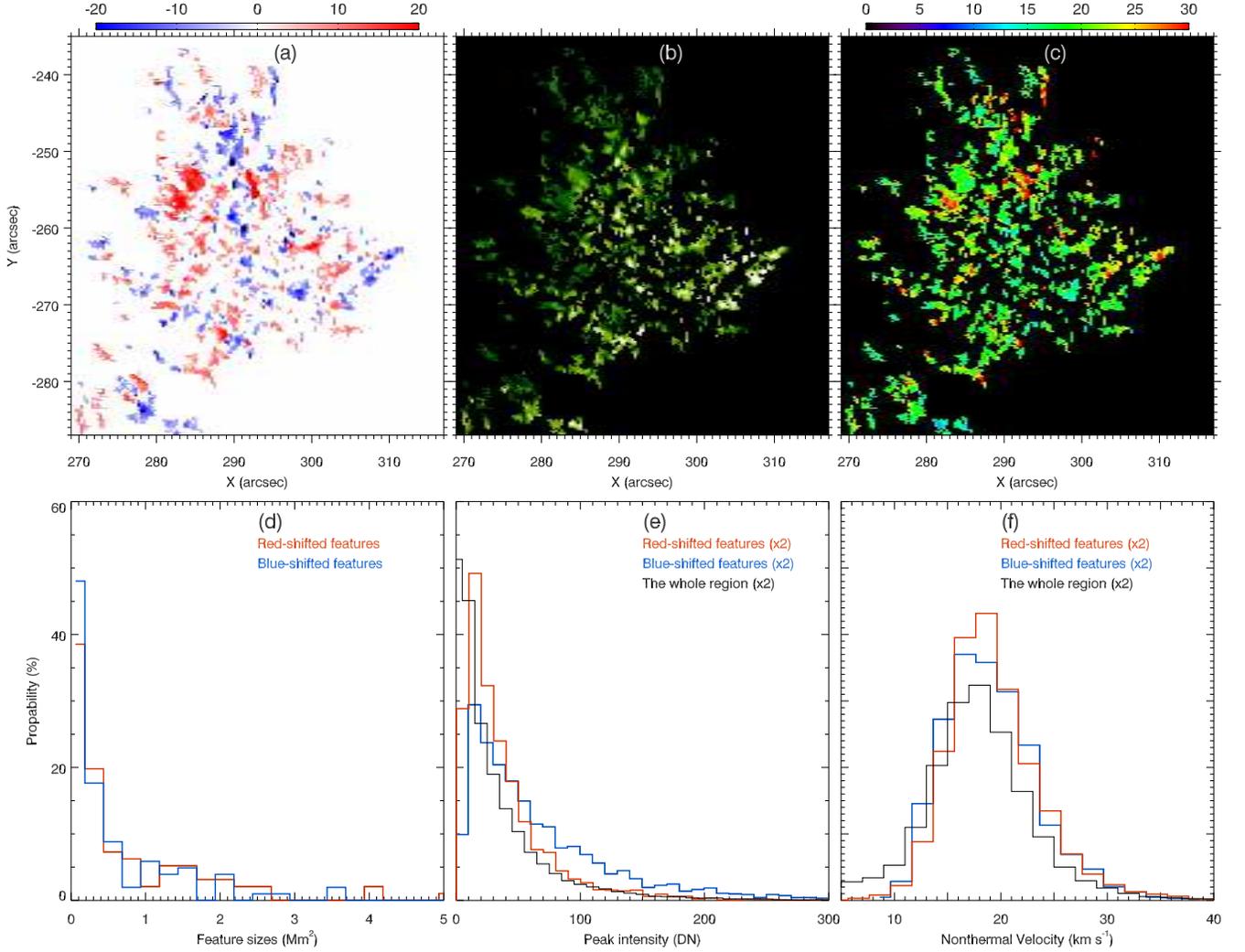}
\caption{The identified velocity structures in the transition region underneath the active region upflows as seen in IRIS \siiv\,1403\,\AA.
(a): the map of Doppler velocity structures; (b): the corresponding intensity map in logarithm scale; (c): the corresponding nonthermal velocity map;
(d): The histograms of the sizes of the identified structures; (e): histograms of the \siiv\,1403\,\AA\ peak intensity for the pixels of the identified blue-shifted features (blue curve), the red-shifted features (red curve) and the whole region (black curve); (f): histograms of the \siiv\,1403\,\AA\ nonthermal velocity for the pixels of the identified blue-shifted features (blue curve), the red-shifted features (red curve) and the whole region (black curve).
The histograms shown in panels e\&f are enhanced by a factor of 2.
\label{fig:vfeature}}
\end{figure*}
%%%%

\begin{figure*}[ht!]
\centering
\includegraphics[width=0.8\textwidth,clip,trim=0cm 1cm 1cm 1cm]{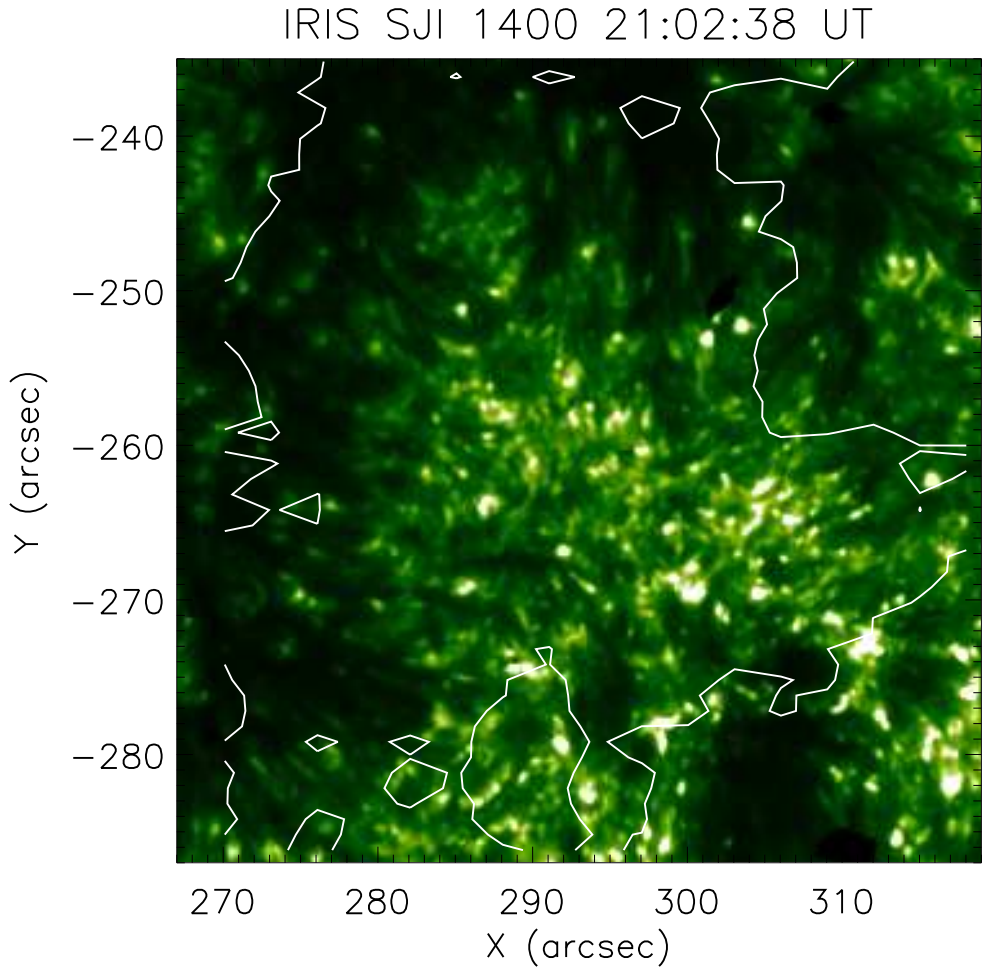}
%\plotone{SJ1400.eps}
\caption{The active region upflows seen in IRIS SJ 1400\,\AA\ passband.
The contours denote the boundaries of the upflow region seen in EIS Fe\,{\sc xii}\,195.12\,\AA\ line (as same as those shown in Figure\,\ref{fig:fov}).
An associated animation is provided online, which displays the evolution of the region from 21:02\,UT to 21:25\,UT.
\label{fig:sj1400}}
\end{figure*}
%%%%

\begin{figure*}[ht!]
\centering
\includegraphics[width=0.8\textwidth,clip,trim=0cm 0cm 0cm 0cm]{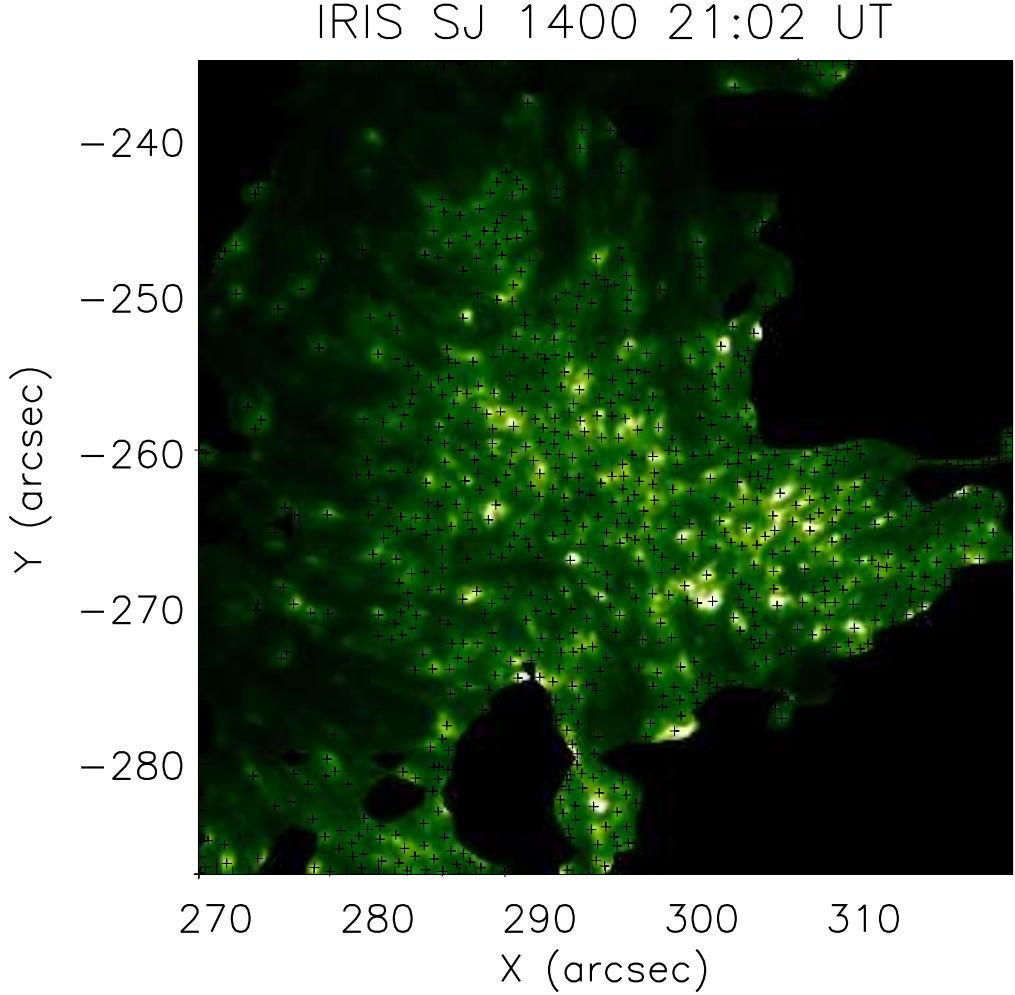}
%\plotone{SJ1400.eps}
\caption{An example frame of IRIS SJ 1400\,\AA\ images with marks of identified transition region dynamic bright dots (plus symbols).
Please note that the marked plus symbols include all the identified bright dots whose living periods cover the observing time of the frame as shown.
\label{fig:swamis_id_exp}}
\end{figure*}
%%%%

\begin{figure*}[ht!]
\centering
\includegraphics[width=\textwidth,clip,trim=0cm 0cm 0cm 0cm]{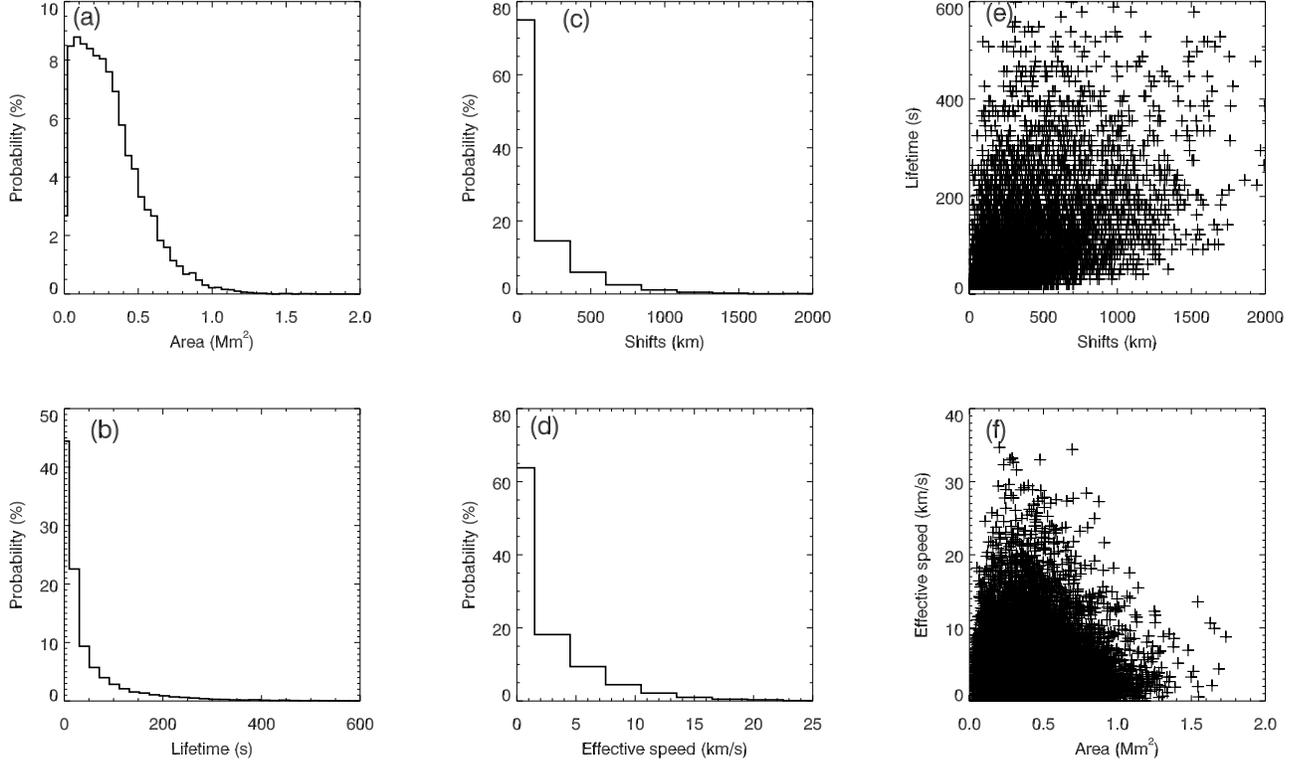}
%\plotone{SJ1400.eps}
\caption{Statistics of parameters of the dynamic bright dots in the transition region underneath the active region upflows as determined from SWAMIS.
(a): the histogram of their areas; (b): the histogram of their lifetimes; (c): the histogram of their shifts from birth to death in the plane of sky; (d): the histogram of their effective speeds; (e): the scatter plot showing the distribution of the dynamics bright dots in the space of shifts vs. lifetimes; (f): the scatter plot showing showing the distribution of the dynamics bright dots in the space of areas vs. 
effective speeds.
\label{fig:swamis_hist}}
\end{figure*}
%%%%

\begin{figure*}[ht!]
\centering
\includegraphics[width=\textwidth,clip,trim=0cm 0cm 12cm 0cm]{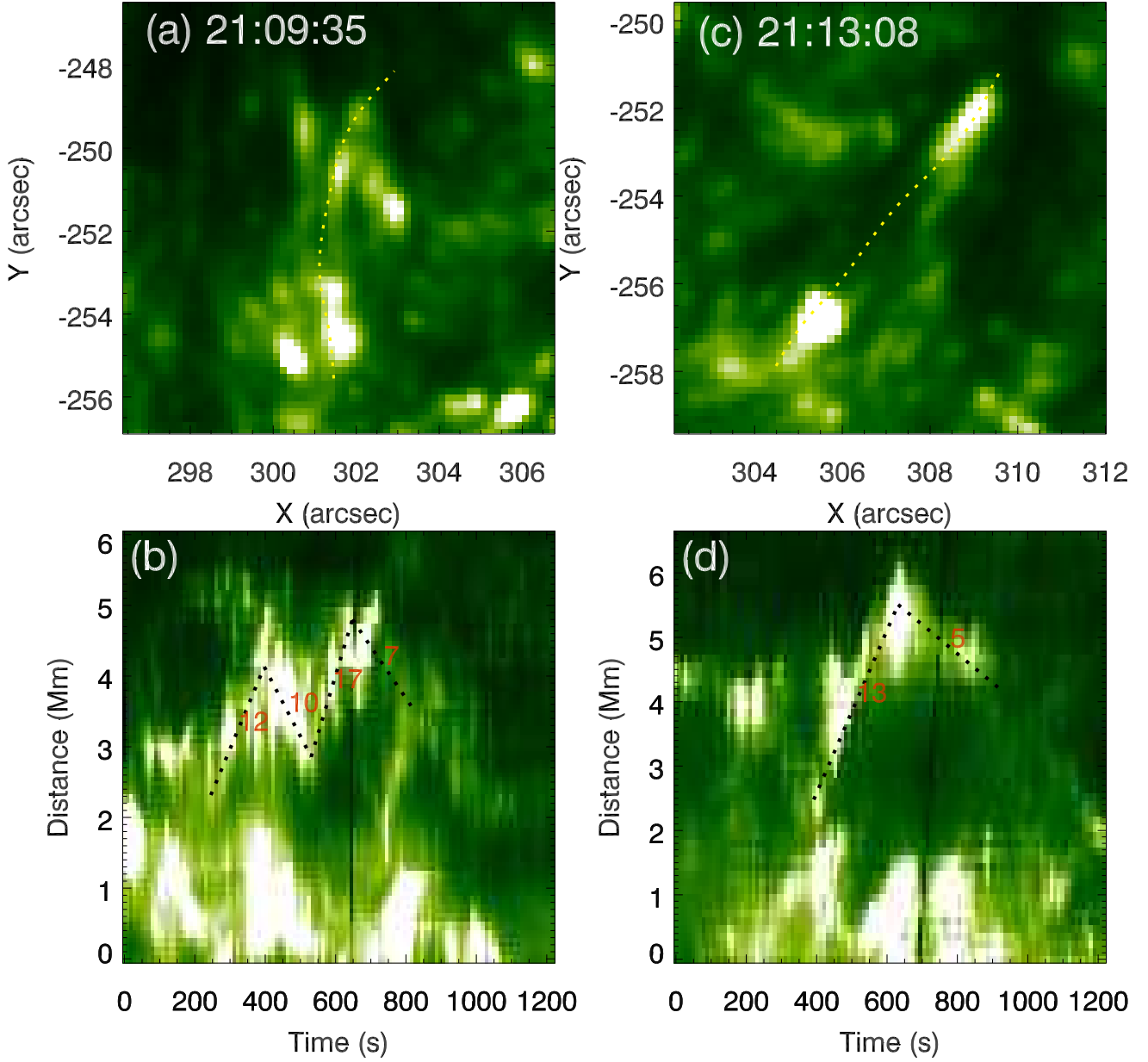}
%\plotone{SJ1400.eps}
\caption{Two examples of transition region surge events appearing at the edge the field of the upflow region.
The top row (panels a\&c) shows the regions of these surges in IRIS 1400\,\AA\ passband, on which the dotted lines denote the trajectories of the surges.
The bottom row (panels b\&d) shows the corresponding time-distance maps obtained along the trajectories of the surges (i.e. the dotted lines shown in the top row).
The dotted lines shown in the bottom row trace the motions of the blobs in the surge and the speeds (in unit of \kms) determined from them are indicated in red.
\label{fig:surges}}
\end{figure*}
%%%%

\begin{figure*}[ht!]
\includegraphics[width=\textwidth,clip,trim=0cm 2cm 0cm 0cm]{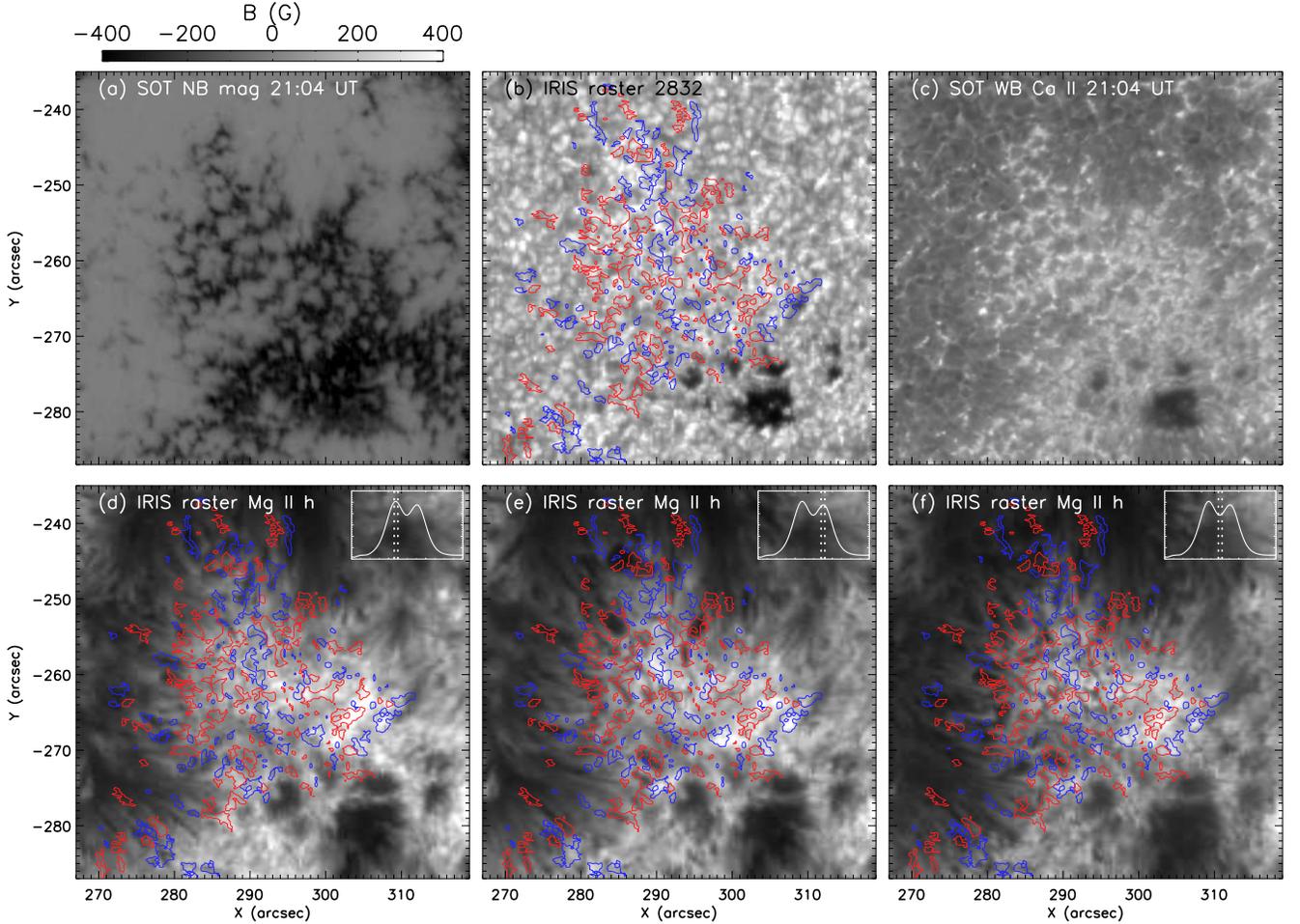}
\caption{The upflow region viewed in SOT line-of-sight magnetogram (panel a), IRIS intensity map summed from 2831.5--2834.0\,\AA\ (panel b), SOT Ca\,{\sc ii}\,h (panel c), and IRIS intensity map at the blue wing (panel d), red wing (panel e) and line center (panel f) of Mg\,{\sc ii}\,h.
The Mg\,{\sc ii}\,h images are obtained from the spectral ranges denoted by the dotted lines on the line profiles shown in their corresponding panel.
In panels b, d--f, the identified blue-shifted and red-shifted features in \siiv\,1403\,\AA\ Doppler map are denoted by the contours in blue and red colour, respectively.\label{fig:sotchro}}
\end{figure*}
%%%%
\section{Observations}\label{sec:obs}
The coordinated data analysed here were taken on 27 December 2013 by IRIS\,\citep{2014SoPh..289.2733D}, the EUV Imaging Spectrometer\,\citep[EIS,][]{2007SoPh..243...19C} and the Solar Optical Telescope\,\citep[SOT,][]{2008SoPh..249..167T} on-board Hinode\,\citep{2007SoPh..243....3K} and the Atomspheric Imaging Assembly\,\citep[AIA,][]{2012SoPh..275...17L} and the Helioseismic and Magnetic Imager\,\citep[HMI,][]{2012SoPh..275..207S} on-board the Solar Dynamics Observatory\,\citep[SDO,][]{2012SoPh..275....3P}.
The observational experiment coordinating IRIS and Hinode was targeting at the active region of NOAA 11934.
The field-of-views (FOVs) of IRIS and EIS on the context image (AIA 171\,\AA) can be seen in Figure\,\ref{fig:fov}a.

\par
The IRIS data studied here were taken from 21:02\,UT to 21:36\,UT on the day, while the instrument was running in a ``very dense raster'' mode.
The IRIS spectrograph slit (with a width of 0.35\arcsec) scanned the region for 400 steps with a step size of 0.35\arcsec and an exposure time of 4\,s.
The pixel size along the slit is 0.17\arcsec.
In this work, we mainly analyse the spectra in the windows of Si\,{\sc iv}\,1403\,\AA, 2832\,\AA\ and Mg\,{\sc ii}\,h, all with a spectral sampling of $\sim25.4$\,m\AA\ pixel$^{-1}$.
The slit-jaw images (SJIs) taken at the 1400\,\AA\ passband with a spatial resolution of 0.33\arcsec and a cadence of 10\,s are included.
The latest version of level-2 data provided by the IRIS team are reanalysed and the wavelength calibration is better than 0.5\,\kms\,\citep{2018SoPh..293..149W}.
To derive the rest wavelength of Si\,{\sc iv} 1403\,\AA, the same procedures as described in \citet{2015ApJ...810...46H} are adopted.
To derive nonthermal velocity, the instrumental broadening (FWHM) is given as 26\,m\AA\,\citep{2014SoPh..289.2733D}.

\par
EIS scanned the region repeatedly with 27 rasters with a 2\arcsec\ slit, a step size of 3\arcsec\ and an exposure time of 3\,s.
It started taking observation from 21:13\,UT and took about 4 minutes to complete one raster.
The data are then calibrated with the standard procedures including the wavelength calibration.
Here, we mainly analysed the data of Fe\,{\sc xii}\,195\,\AA, which provides the general morphology of the active region upflows, including its location and size.
The rest wavelength of the Fe\,{\sc xii} line was obtained from a relative quiet region at the top right of the EIS FOV and given as 195.119\,\AA.

\par
SOT provides three magnetograms taken at Na\,{\i} 5896\,\AA\ and 11 snapshots taken at Ca\,{\sc ii}\,H.
The spatial resolutions of the magnetograms and the Ca\,{\sc ii}\,H images are 0.32\arcsec and 0.22\arcsec, respectively.
Although these data do not allow a study on the evolution of the region, they can show the magnetic and chromospheric structures in a great detail.

\par
Both AIA images and HMI line-of-sight magnetograms are used to co-aligned the data from different instruments and also give a global overview of the connection of the region.
Both data have a spatial resolution of 1.2\arcsec.
The AIA 171\,\AA\ images have a cadence of 12\,s and the HMI magnetograms have a cadence of 45\,s.

\section{Data analyses and results}\label{sec:res}
The region of coronal upflows shows a blue shift of 10--20\,\kms\ in Fe\,{\sc xii} (with a formation temperature of $1.6\times10^6$\,K), locating at the east side of the active region (see Figure\,\ref{fig:fov}a--c).
Here we focus on the core of the upflows, which presents a dark region with roots of fan loops surrounded as seen in AIA 171\,\AA\ images (see Figure\,\ref{fig:fov}a).
Such morphology is typical for active region upflows as shown in previous studies\,\citep[e.g.][]{2008ApJ...676L.147H,2020ApJ...894..144B,2020ApJ...903...68P}.
The blue-shifted area of the selected region as seen in EIS Fe\,{\sc xii}\,195\,\AA\ line  (see the contours shown in Figure\,\ref{fig:fov}c) is about 897\,mega-meter$^2$ ($Mm^2$).
If we consider only the dark region surrounded by bright footpoints of the fan loops, it has an area of about 156\,$Mm^2$ (see the region between the two dashed lines in Figure\,\ref{fig:fov}a).

\subsection{Analyses of \siiv\,1403\,\AA\ spectral data}
In IRIS Si\,{\sc iv}\,1403\,\AA\ (with a formation temperature of $8\times10^4$\,K) observations, the region is highly structured, where abundant dot-like brightenings are seen on the intensity map (Figure\,\ref{fig:fov}e\&h) and very fine blue- and red-shifted features on the Doppler velocity map (Figure\,\ref{fig:fov}f\&i).
Comparing to a surrounding region with similar intensity structures (see the region around X=[310,340], Y=[-270,-240] in Figure\,\ref{fig:fov}f), it is notable that the upflow region includes more blue-shifted structures.
This is in agreement with the results shown in \citet{2020ApJ...903...68P}.
The nonthermal velocity map of the region shows no significant difference from the surrounding area (see Figure\,\ref{fig:fov}d\&g), which is similar to that of a normal plage region\,\citep[e.g.][]{2016AIPC.1720b0001H}.
\par
We then manually identify the patches of fine structures in the \siiv\,1403\,\AA\ Doppler map underneath the coronal upflows (i.e. blue shifts in Fe\,{\sc xii} line, see the contour in Figure\,\ref{fig:fov}).
The edge of each (blue-)red-shifted patch is defined at the place where the absolute speeds are less than 5\,\kms.
Most of the identified velocity patches have Doppler shifts from 10 to 20\,\kms.
The identified small-scaled velocity structures and their corresponding intensity and nonthermal velocity ones are shown in Figure\,\ref{fig:vfeature}.
In total, 102 blue-shifted patches and 96 red-shifted patches are identified.
The sizes of the blue-shifted patches are in the range of 0.05--6.79\,$Mm^2$, and those of the red-shifted patches are ranging from 0.05--12.18\,$Mm^2$.
With these samples, 99\% of the blue-shifted ones and 97\% of the red-shifted ones are smaller than 5\,$Mm^2$, in which more than 70\% are smaller than 1\,$Mm^2$ (see Figure\,\ref{fig:fov}d).
The total areas of the blue-shifted and red-shifted patches are 74\,$Mm^2$ and 98\,$Mm^2$, respectively.
Please notice that the average Doppler velocity in this region remains red-shifted.
We can see that the total area of the upflow region in the corona is larger than the size of any single patch of these velocity structures in the transition region for 2--3 magnitudes.
The histogram of \siiv\ intensity of blue-shifted features shows a clear enhancement at higher intensity (see Figure\,\ref{fig:fov}e).
The average intensity of the blue-shifted pixels is 75\,DN in comparison to 43\,DN for the red-shifted ones. 
This indicates that comparing to the red-shifted features, the blue-shifted ones tend to be associated with brighter regions in the transition region.
The nonthermal velocities, however, do not show any obvious difference in their histograms (Figure\,\ref{fig:vfeature}f),
suggesting that heating in these velocity structures should be statistically similar to the ambient region.

\subsection{Transition region dynamics seen in IRIS 1400\,\AA\ slit-jaw images}
From IRIS SJ 1400\,\AA\ observations (Figure\,\ref{fig:sj1400} and the associated animation), we can see the upflow region is full of dynamic dot-like brightenings.
Some of them are corresponding to the bright dots in the \siiv\ intensity image.

\par
In order to study the dynamic properties of these bright dots, we adopt a feature tracking method, the ``Southwest Automatic Magnetic Identification Suite'' (or ``SWAMIS'' in short), to identify and trace them in the IRIS SJ 1400\,\AA\ observations.
SWAMIS has been widely used in tracking magnetic features on the photosphere\,\citep[e.g.][etc.]{2008PhDT........20L,2009ApJ...698...75P,2012A&A...548A..62H,2014ApJ...788....7L}, and the detailed methodology of SWAMIS can be seen in a series of papers\,\citep{2007ApJ...666..576D,2008ApJ...674..520L,2010ApJ...720.1405L,2013ApJ...774..127L}.
To setup the program, 
the high threshold was set to be $2\sigma$ above the average of the intensities of all pixels of the upflow region in the entire observing period, which triggers identification of a feature;
and the low threshold was set to be the average value, which involves in the definition of the starting and ending time and also the edges of a feature;
the ``downhill'' method was chosen to detect the edges of a feature, which define the edges of a feature at the zero gradient toward the zero intensity unless reaching the low threshold.
The thresholds were selected in the way that they are given robustly and they can cover most of the bright dots as seen manually (see an example in Figure\,\ref{fig:swamis_id_exp}).
In the observations from 21:02\,UT to 21:25\,UT, SWAMIS identified 20\,667 bright dots in the transition region underneath the coronal upflows.
With the above settings, the area, birth and death time, and birth and death locations of the identified bright dots can be achieved.
In Figure\,\ref{fig:swamis_hist}, we give the statistics of the areas (panel a), lifetimes (panel b), shifts between birth location and death location (panel c) and effective speeds (i.e. shift divided by lifetime, panel d) and distributions of the bright dots in the spaces of shifts vs. lifetimes (panel e) and areas vs. effective speeds (panel f).

\par
The areas of all these bright dots have an average of 0.3\,$Mm^2$.
The histogram of the areas peaks at about 0.1\,$Mm^2$ and 99\% of the bright dots are smaller than 1\,$Mm^2$.
These again confirm the transition region dynamics underneath the upflow region occurs in very fine scale.
While about 95\% of the bright dots have lifetimes smaller than 200\,s,
we also notice that about 44\% are identified in only one frame of the images, indicating their lifetimes are smaller than 10\,s.
Quantitatively, this confirms a highly dynamic nature of the transition region underneath the upflow region.
We also confirm that many bright dots can be born in the same locations, suggesting that they are intermittent phenomena.

\par
From the animation of SJ\,1400\,\AA\ observations (Figure\,\ref{fig:sj1400}), we can see that most of the bright dots at the center of the upflow region are mostly localized without significant motions away from the initial locations.
This is confirmed by the measurement of shifts of their brightest centers given by SWAMIS (Figure\,\ref{fig:swamis_hist}c), where we can see more than 75\% of the bright dots hardly have any movement and more than 98\% shift for a distance less than 1\,Mm.
If we took into account their sizes, only about 6\% can shift further than their initial boundaries.
Accordingly, the effective speeds of about 65\% of the bright dots are around zero and 98\% are less than 15\,\kms.

\par
The scatter plot in the panel (e) of Figure\,\ref{fig:swamis_hist} shows a weak trend in statistics that bright dots with larger shifts tend to have longer lifetimes, although the distribution is rather disperse.
The scatter distribution of the bright dots in the space of areas vs. effective speeds (panel (f) of Figure\,\ref{fig:swamis_hist}) indicates that high speed ($>20$\,\kms) bright dots are concentrate at areas around 0.3\,$Mm^2$ and bright dots with larger areas tend to have a smaller upper limit of speeds.

\par
Surge-like motions can be seen in the bright dots at the edge of the region, where elongated dark threads are also abundant (suggesting inclined magnetic field).
In Figure\,\ref{fig:surges}, we show two examples of surges.
A distinguished characteristics is that these surges normally show a bright front (see the top row of Figure\,\ref{fig:surges}),
which is similar to light walls occurring above light bridges of sunspots\,\citep[see e.g.][]{2015ApJ...804L..27Y,2016A&A...589L...7H}, and might be a hint of shocks\,\citep{2017ApJ...838....2Z,2017ApJ...848L...9H,2018ApJ...855...65H}.
The upward motions of these surges have speeds of 12--17\,\kms, and the downward speeds are slightly smaller (the bottom panel of Figure\,\ref{fig:surges}).
These values are consistent with the velocity structures measured in the spectral data.
Therefore, it remains possible that a part of bright dots in the center region also have surge-like motions but they cannot be resolved due to the projection effect.

\subsection{The chromosphere and photosphere underneath the active region upflows}
In Figure\,\ref{fig:sotchro}, we further investigate the counterpart of the upflow region in the lower chromosphere and photosphere.
The high resolution magnetogram shows that the region is dominant by negative polarity (Figure\,\ref{fig:sotchro}a).
This suggests (quasi-)open magnetic field could be exist in a part of this region, in agreement with previous studies\,\citep[e.g.][]{2009ApJ...705..926B}. 
The radiation image near 2832\,\AA\ shows that the upflow region has a typical morphology of faculae (Figure\,\ref{fig:sotchro}b).
Consistent with the faculae in the photosphere, chromospheric plage in the 
region is clearly seen in the SOT Ca\,{\sc ii}\,h image (Figure\,\ref{fig:sotchro}c).
%Although some transition region velocity structures locate above bright granule cells,
%dark granular lanes seem to be more prominent in the locations of the blue-shifted features.

\par
The upflow region seen in Mg\,{\sc ii} line is shown in Figure\,\ref{fig:sotchro}d--f.
We can see that the intensity maps of Mg\,{\sc ii}\,h wings and line center are generally in agreement with each other, suggesting this region in the middle and upper chromosphere are coherent\,\citep{2013ApJ...772...90L,2013ApJ...778..143P}.
At the edge of the region, the elongated dark threads shown in SJ 1400\,\AA\ images are also seen in the chromosphere.
While cross-checking these maps in detail, we can see that some blue-shifted structures present as dark features in the blue wing image but as bright ones in the red wing, or vice-versa for some red-shifted structures.
This suggests that plasma flows in the chromosphere and transition region are coherent in these locations\,\citep[see relevant discussion in][]{2020ApJ...889..124T}.
This is consistent with previous studies as summarized in the introduction.

\section{Summary and discussion}\label{sec:con}
To infer at what scale the transition region connects to the coronal upflows at the edge of active region, in the present study we investigate the statistical properties of small-scale dynamics in the transition region underneath the upflows of active region NOAA 11934.
Consistent with previous studies, the transition region underneath the coronal upflows is also highly structured.
While the upflows in the corona are a continuous region with an area of about 897\,$Mm^2$, the velocity structures in the transition region are finely scaled as shown on the \siiv\,1403\,\AA\ Doppler map.
In the \siiv\,1403\,\AA\ Doppler velocity map, we identified 102 blue-shifted patches and 96 red-shifted patches.
We found that 99\% of the blue-shifted patches and 97\% of the red-shifted patches are smaller than 5\,$Mm^2$, in which more than 70\% are smaller than 1\,$Mm^2$.
The total area of the blue-shifted features is about 74\,$Mm^2$, a magnitude smaller than the coronal upflows.
Statistically, the \siiv\,1403\,\AA\ intensities of the blue-shifted features are biased towards the large values, comparing to those of the red-shifted ones.
The histograms of nonthermal velocities of both the blue- and red-shifted features do not show any obvious difference from that of the entire region.

\par
The IRIS SJ 1400\,\AA\ images show that dynamic bright dots spread all over the transition region underneath the coronal upflow region.
Using SWAMIS, more than 20\,000 bright dots in the SJ 1400\,\AA\ images were identified and tracked.
We found that their average area is 0.3\,$Mm^2$.
The distribution of their areas peaks at 0.1\,$Mm^2$ and 99\% of them are smaller than 1\,$Mm^2$.
Their lifetimes are from 10\,s to a few minutes, in which more than 95\% are less than 200\,s. 
Statistically, about 94\% of the bright dots do not show any distinguishable shifts in the plane of sky.
Some bright dots might have high apparent speeds ($>$20\,\kms), and the distribution of their areas concentrates at around 0.3\,$Mm^2$.
Surge-like motions with speeds of 10--20\,\kms\ are clearly seen at the edge of the region, and they normally have a bright front.
We speculate that similar motions should also exist in the center part of the region, but they cannot be resolved due to the projection effect.

\par
The high-resolution magnetogram reveals that the upflow region is of single polarity, indicating a (quasi-)open magnetic geometry.
The coronal upflows locate above faculae as seen in IRIS 2832\,\AA\ image and plage region in the SOT Ca\,{\sc ii} image, suggesting an active magnetic environment.
In agreement with the previous studies (see summary in the introduction),
we found that plasma flows in some velocity structures in \siiv\,1403\,\AA\ are consistent with those revealed by \mgii\,h, suggesting a coherence of plasma flows in these places between the chromosphere and transition region.

\par
In light of these observations, we might further discuss on the formation of the active region upflows in the corona.
As summarised in the introduction, many previous studies found that active region upflows have strong connection to dynamics of the transition region or even lower.
This is also confirmed by our observations.
In the present study, we further carry out a statistic analysis to small-scale dynamics in the transition region underneath the active region upflows.
The size and dynamic properties of these structures indicate that the connection between the transition region and the coronal upflows is most likely taken place in small scale ($\lesssim$1\,$Mm^2$) and a rapid and intermittent process (evolving in a time scale less than a few minutes).
The question is how such small-scale dynamics in the transition region links to the upflows in the corona.

\par
An interpretation is that the plasma in the transition region is directly heated and pumped into the corona.
In this way, the coronal upflows should be structured in smaller scale rather than a continuous region, since the upflows in the transition region are localised with sizes of about 1\,$Mm^2$.
Such smaller structures might not be able to resolve with EIS data, and spectral data with higher resolutions from such as SPICE\,\citep{2020A&A...642A..14S} aboard the Solar Orbiter\,\citep[SolO,][]{2020A&A...642A...1M} should be helpful.
Comparing the total blue-shifted area in the transition region and the coronal upflows, we can estimate a filling factor of about 0.08 for the EIS observations of this coronal upflow region.
For EIS observations of typical fan loops, \citet{2012ApJ...744...14Y} found a filling factor of about 0.2,
which is more than twice of the value derived here.
A possible reason for this inconsistency could be the expansion of the fan loops and their projection effects.
For an isolated typical fan loop shown in the south of the studied field-of-view (see the region denoted by a trapezoid in Figure\,\ref{fig:fov}a), we estimate from the AIA\,171\,\AA\ image that a fan with a cross-section of about 6\arcsec\ near the footpoint can expand to about 20\arcsec\ at a distance of 25\arcsec\ on the projection plate.
Suppose the fan has a circular disc of the footpoint (with a diameter of 6\arcsec), we found that the projection area can be more than 10 times of the footpoint.
Although this factor can be seriously influenced by inclined angles and overlapping effects, the expansion of fan loops can easily explain the relatively small filling factor estimated here.
Another reason could be the underestimation of the area of upflows in the transition region, since only one raster of data was taken here and one cannot exclude possible contribution of regions other than the blue-shifted patches.

\par
Alternatively, the small scale dynamics in the transition region can just act as drivers that stimulate the coronal base of the open field region and then force the plasmas there moving upward to form the coronal upflows.
Similarly, evidence that dynamics from the lower solar atmosphere can drive coronal propagating disturbances has been reported by some studies\,\citep[e.g.][etc.]{2015ApJ...807...71P,2015ApJ...809L..17J,2018ApJ...855...65H}.
With this scenario, the upflows in the transition region can fall back after stimulating the coronal base and transferring energy to the corona, thus it explains why the blue-shifted features tend to be brighter than the red-shifted ones.
Also, the transition region dynamics can have any sizes and not be necessary to be comparable to the coronal upflows.
The abundance of the small-scale dynamics in the transition region at any given time should be able to provide enough drivers to the coronal base.
We might assume a funnel geometry of the upflows as similar to those in coronal holes\,\citep[see e.g.][]{1986SoPh..105...35D,2005Sci...308..519T}, then two possible processes might involve in driving the coronal upflows.
One is interaction with shocks, in which the shock energy carried by surges hit the coronal base of the funnel and transfer energy to corona\,\citep{2018ApJ...855...65H}.
Another one is interchange reconnection between a funnel and small-scale loops\,\citep{2005ApJ...626..563F}.
Actually, interchange reconnection scenario for active region upflows has been proposed or suggested by many previous studies\,\citep[e.g.][etc.]{2009ApJ...705..926B,2011A&A...526A.137D,2021arXiv210410234B}.
The small-scale loops here could be understood as small transition region loops that might be propelled by surges and then interact with coronal funnels, or small magnetic islands associated with energetic events in the transition region\,\citep{2015ApJ...813...86I} that carry close fields moving and interacting with coronal funnels.

\par
{\it Acknowledgments:}
We are grateful to the critical and constructive comments from the anonymous reviewer that help improve a lot the work.
This research is supported by National Natural Science Foundation of China (41974201,U1831112) and the Young Scholar Program of Shandong University, Weihai (2017WHWLJH07). 
IRIS is a NASA small explorer mission developed and operated by LMSAL with mission operations executed at NASA Ames Research center and major contributions to downlink communications funded by ESA and the Norwegian Space Centre.
Hinode is a Japanese mission developed and launched by ISAS/JAXA, collaborating with NAOJ as a domestic partner, NASA and STFC (UK) as international partners.
Scientific operation of the Hinode mission is conducted by the Hinode science team organized at ISAS/JAXA.
This team mainly consists of scientists from institutes in the partner countries.
Support for the post-launch operation is provided by JAXA and NAOJ(Japan), STFC (U.K.), NASA, ESA, and NSC (Norway).
The AIA and HMI data are used by courtesy of NASA/SDO, the AIA and HMI teams and JSOC.
SWAMIS is a freely available open source feature tracking suite written by Craig DeForest and Derek Lamb at the Southwest Research Institute Department of Space Studies in Boulder, Colorado.
\bibliography{TRflows}{}

\begin{thebibliography}{}
\expandafter\ifx\csname natexlab\endcsname\relax\def\natexlab#1{#1}\fi
\providecommand{\url}[1]{\href{#1}{#1}}
\providecommand{\dodoi}[1]{doi:~\href{http://doi.org/#1}{\nolinkurl{#1}}}
\providecommand{\doeprint}[1]{\href{http://ascl.net/#1}{\nolinkurl{http://ascl.net/#1}}}
\providecommand{\doarXiv}[1]{\href{https://arxiv.org/abs/#1}{\nolinkurl{https://arxiv.org/abs/#1}}}

\bibitem[{{Abbo} {et~al.}(2016){Abbo}, {Ofman}, {Antiochos}, {Hansteen},
  {Harra}, {Ko}, {Lapenta}, {Li}, {Riley}, {Strachan}, {von Steiger}, \&
  {Wang}}]{2016SSRv..201...55A}
{Abbo}, L., {Ofman}, L., {Antiochos}, S.~K., {et~al.} 2016, \ssr, 201, 55,
  \dodoi{10.1007/s11214-016-0264-1}

\bibitem[{{Baker} {et~al.}(2017){Baker}, {Janvier}, {D{\'e}moulin}, \&
  {Mandrini}}]{2017SoPh..292...46B}
{Baker}, D., {Janvier}, M., {D{\'e}moulin}, P., \& {Mandrini}, C.~H. 2017,
  \solphys, 292, 46, \dodoi{10.1007/s11207-017-1072-9}

\bibitem[{{Baker} {et~al.}(2009){Baker}, {van Driel-Gesztelyi}, {Mandrini},
  {D{\'e}moulin}, \& {Murray}}]{2009ApJ...705..926B}
{Baker}, D., {van Driel-Gesztelyi}, L., {Mandrini}, C.~H., {D{\'e}moulin}, P.,
  \& {Murray}, M.~J. 2009, \apj, 705, 926, \dodoi{10.1088/0004-637X/705/1/926}

\bibitem[{{Barczynski} {et~al.}(2021){Barczynski}, {Harra}, {Kleint}, {Panos},
  \& {Brooks}}]{2021arXiv210410234B}
{Barczynski}, K., {Harra}, L., {Kleint}, L., {Panos}, B., \& {Brooks}, D.~H.
  2021, arXiv e-prints, arXiv:2104.10234.
\newblock \doarXiv{2104.10234}

\bibitem[{{Boutry} {et~al.}(2012){Boutry}, {Buchlin}, {Vial}, \&
  {R{\'e}gnier}}]{2012ApJ...752...13B}
{Boutry}, C., {Buchlin}, E., {Vial}, J.~C., \& {R{\'e}gnier}, S. 2012, \apj,
  752, 13, \dodoi{10.1088/0004-637X/752/1/13}

\bibitem[{{Brooks} {et~al.}(2021){Brooks}, {Harra}, {Bale}, {Barczynski},
  {Mandrini}, {Polito}, \& {Warren}}]{2021arXiv210603318B}
{Brooks}, D.~H., {Harra}, L., {Bale}, S.~D., {et~al.} 2021, arXiv e-prints,
  arXiv:2106.03318.
\newblock \doarXiv{2106.03318}

\bibitem[{{Brooks} {et~al.}(2015){Brooks}, {Ugarte-Urra}, \&
  {Warren}}]{2015NatCo...6.5947B}
{Brooks}, D.~H., {Ugarte-Urra}, I., \& {Warren}, H.~P. 2015, Nature
  Communications, 6, 5947, \dodoi{10.1038/ncomms6947}

\bibitem[{{Brooks} \& {Warren}(2011)}]{2011ApJ...727L..13B}
{Brooks}, D.~H., \& {Warren}, H.~P. 2011, \apjl, 727, L13,
  \dodoi{10.1088/2041-8205/727/1/L13}

\bibitem[{{Brooks} \& {Warren}(2012)}]{2012ApJ...760L...5B}
---. 2012, \apjl, 760, L5, \dodoi{10.1088/2041-8205/760/1/L5}

\bibitem[{{Brooks} {et~al.}(2020){Brooks}, {Winebarger}, {Savage}, {Warren},
  {De Pontieu}, {Peter}, {Cirtain}, {Golub}, {Kobayashi}, {McIntosh},
  {McKenzie}, {Morton}, {Rachmeler}, {Testa}, {Tiwari}, \&
  {Walsh}}]{2020ApJ...894..144B}
{Brooks}, D.~H., {Winebarger}, A.~R., {Savage}, S., {et~al.} 2020, \apj, 894,
  144, \dodoi{10.3847/1538-4357/ab8a4c}

\bibitem[{{Chen} {et~al.}(2011){Chen}, {Ding}, {Chen}, \&
  {Harra}}]{2011ApJ...740..116C}
{Chen}, F., {Ding}, M.~D., {Chen}, P.~F., \& {Harra}, L.~K. 2011, \apj, 740,
  116, \dodoi{10.1088/0004-637X/740/2/116}

\bibitem[{{Culhane} {et~al.}(2007){Culhane}, {Harra}, {James}, {Al-Janabi},
  {Bradley}, {Chaudry}, {Rees}, {Tandy}, {Thomas}, {Whillock}, {Winter},
  {Doschek}, {Korendyke}, {Brown}, {Myers}, {Mariska}, {Seely}, {Lang}, {Kent},
  {Shaughnessy}, {Young}, {Simnett}, {Castelli}, {Mahmoud}, {Mapson-Menard},
  {Probyn}, {Thomas}, {Davila}, {Dere}, {Windt}, {Shea}, {Hagood}, {Moye},
  {Hara}, {Watanabe}, {Matsuzaki}, {Kosugi}, {Hansteen}, \&
  {Wikstol}}]{2007SoPh..243...19C}
{Culhane}, J.~L., {Harra}, L.~K., {James}, A.~M., {et~al.} 2007, \solphys, 243,
  19, \dodoi{10.1007/s01007-007-0293-1}

\bibitem[{{Culhane} {et~al.}(2014){Culhane}, {Brooks}, {van Driel-Gesztelyi},
  {D{\'e}moulin}, {Baker}, {DeRosa}, {Mandrini}, {Zhao}, \&
  {Zurbuchen}}]{2014SoPh..289.3799C}
{Culhane}, J.~L., {Brooks}, D.~H., {van Driel-Gesztelyi}, L., {et~al.} 2014,
  \solphys, 289, 3799, \dodoi{10.1007/s11207-014-0551-5}

\bibitem[{{De Pontieu} \& {McIntosh}(2010)}]{2010ApJ...722.1013D}
{De Pontieu}, B., \& {McIntosh}, S.~W. 2010, \apj, 722, 1013,
  \dodoi{10.1088/0004-637X/722/2/1013}

\bibitem[{{De Pontieu} {et~al.}(2014){De Pontieu}, {Title}, {Lemen}, {Kushner},
  {Akin}, {Allard}, {Berger}, {Boerner}, {Cheung}, {Chou}, {Drake}, {Duncan},
  {Freeland}, {Heyman}, {Hoffman}, {Hurlburt}, {Lindgren}, {Mathur}, {Rehse},
  {Sabolish}, {Seguin}, {Schrijver}, {Tarbell}, {W{\"u}lser}, {Wolfson},
  {Yanari}, {Mudge}, {Nguyen-Phuc}, {Timmons}, {van Bezooijen}, {Weingrod},
  {Brookner}, {Butcher}, {Dougherty}, {Eder}, {Knagenhjelm}, {Larsen},
  {Mansir}, {Phan}, {Boyle}, {Cheimets}, {DeLuca}, {Golub}, {Gates}, {Hertz},
  {McKillop}, {Park}, {Perry}, {Podgorski}, {Reeves}, {Saar}, {Testa}, {Tian},
  {Weber}, {Dunn}, {Eccles}, {Jaeggli}, {Kankelborg}, {Mashburn}, {Pust},
  {Springer}, {Carvalho}, {Kleint}, {Marmie}, {Mazmanian}, {Pereira}, {Sawyer},
  {Strong}, {Worden}, {Carlsson}, {Hansteen}, {Leenaarts}, {Wiesmann},
  {Aloise}, {Chu}, {Bush}, {Scherrer}, {Brekke}, {Martinez-Sykora}, {Lites},
  {McIntosh}, {Uitenbroek}, {Okamoto}, {Gummin}, {Auker}, {Jerram}, {Pool}, \&
  {Waltham}}]{2014SoPh..289.2733D}
{De Pontieu}, B., {Title}, A.~M., {Lemen}, J.~R., {et~al.} 2014, \solphys, 289,
  2733, \dodoi{10.1007/s11207-014-0485-y}

\bibitem[{{DeForest} {et~al.}(2007){DeForest}, {Hagenaar}, {Lamb}, {Parnell},
  \& {Welsch}}]{2007ApJ...666..576D}
{DeForest}, C.~E., {Hagenaar}, H.~J., {Lamb}, D.~A., {Parnell}, C.~E., \&
  {Welsch}, B.~T. 2007, \apj, 666, 576, \dodoi{10.1086/518994}

\bibitem[{{Del Zanna} {et~al.}(2011){Del Zanna}, {Aulanier}, {Klein}, \&
  {T{\"o}r{\"o}k}}]{2011A&A...526A.137D}
{Del Zanna}, G., {Aulanier}, G., {Klein}, K.~L., \& {T{\"o}r{\"o}k}, T. 2011,
  \aap, 526, A137, \dodoi{10.1051/0004-6361/201015231}

\bibitem[{{D{\'e}moulin} {et~al.}(2013){D{\'e}moulin}, {Baker}, {Mandrini}, \&
  {van Driel-Gesztelyi}}]{2013SoPh..283..341D}
{D{\'e}moulin}, P., {Baker}, D., {Mandrini}, C.~H., \& {van Driel-Gesztelyi},
  L. 2013, \solphys, 283, 341, \dodoi{10.1007/s11207-013-0234-7}

\bibitem[{{Doschek} {et~al.}(2008){Doschek}, {Warren}, {Mariska}, {Muglach},
  {Culhane}, {Hara}, \& {Watanabe}}]{2008ApJ...686.1362D}
{Doschek}, G.~A., {Warren}, H.~P., {Mariska}, J.~T., {et~al.} 2008, \apj, 686,
  1362, \dodoi{10.1086/591724}

\bibitem[{{Dowdy} {et~al.}(1986){Dowdy}, {Rabin}, \&
  {Moore}}]{1986SoPh..105...35D}
{Dowdy}, J.~F., J., {Rabin}, D., \& {Moore}, R.~L. 1986, \solphys, 105, 35,
  \dodoi{10.1007/BF00156374}

\bibitem[{{Edwards} {et~al.}(2016){Edwards}, {Parnell}, {Harra}, {Culhane}, \&
  {Brooks}}]{2016SoPh..291..117E}
{Edwards}, S.~J., {Parnell}, C.~E., {Harra}, L.~K., {Culhane}, J.~L., \&
  {Brooks}, D.~H. 2016, \solphys, 291, 117, \dodoi{10.1007/s11207-015-0807-8}

\bibitem[{{Fisk}(2005)}]{2005ApJ...626..563F}
{Fisk}, L.~A. 2005, \apj, 626, 563, \dodoi{10.1086/429957}

\bibitem[{{Fu} {et~al.}(2015){Fu}, {Li}, {Li}, {Huang}, {Mou}, {Jiao}, \&
  {Xia}}]{2015SoPh..290.1399F}
{Fu}, H., {Li}, B., {Li}, X., {et~al.} 2015, \solphys, 290, 1399,
  \dodoi{10.1007/s11207-015-0689-9}

\bibitem[{{Fu} {et~al.}(2017){Fu}, {Madjarska}, {Xia}, {Li}, {Huang}, \&
  {Wangguan}}]{2017ApJ...836..169F}
{Fu}, H., {Madjarska}, M.~S., {Xia}, L., {et~al.} 2017, \apj, 836, 169,
  \dodoi{10.3847/1538-4357/aa5cba}

\bibitem[{{Galsgaard} {et~al.}(2015){Galsgaard}, {Madjarska}, {Vanninathan},
  {Huang}, \& {Presmann}}]{2015A&A...584A..39G}
{Galsgaard}, K., {Madjarska}, M.~S., {Vanninathan}, K., {Huang}, Z., \&
  {Presmann}, M. 2015, \aap, 584, A39, \dodoi{10.1051/0004-6361/201526339}

\bibitem[{{Guo} {et~al.}(2010){Guo}, {Tian}, \& {He}}]{2010RAA....10.1307G}
{Guo}, L.-J., {Tian}, H., \& {He}, J.-S. 2010, Research in Astronomy and
  Astrophysics, 10, 1307, \dodoi{10.1088/1674-4527/10/12/011}

\bibitem[{{Harra} {et~al.}(2008){Harra}, {Sakao}, {Mandrini}, {Hara}, {Imada},
  {Young}, {van Driel-Gesztelyi}, \& {Baker}}]{2008ApJ...676L.147H}
{Harra}, L.~K., {Sakao}, T., {Mandrini}, C.~H., {et~al.} 2008, \apjl, 676,
  L147, \dodoi{10.1086/587485}

\bibitem[{{Harra} {et~al.}(2017){Harra}, {Ugarte-Urra}, {De Rosa}, {Mandrini},
  {van Driel-Gesztelyi}, {Baker}, {Culhane}, \&
  {D{\'e}moulin}}]{2017PASJ...69...47H}
{Harra}, L.~K., {Ugarte-Urra}, I., {De Rosa}, M., {et~al.} 2017, \pasj, 69, 47,
  \dodoi{10.1093/pasj/psx021}

\bibitem[{{He} {et~al.}(2010){He}, {Marsch}, {Tu}, {Guo}, \&
  {Tian}}]{2010A&A...516A..14H}
{He}, J.~S., {Marsch}, E., {Tu}, C.~Y., {Guo}, L.~J., \& {Tian}, H. 2010, \aap,
  516, A14, \dodoi{10.1051/0004-6361/200913712}

\bibitem[{{Hinode Review Team} {et~al.}(2019){Hinode Review Team}, {Al-Janabi},
  {Antolin}, {Baker}, {Bellot Rubio}, {Bradley}, {Brooks}, {Centeno},
  {Culhane}, {Del Zanna}, {Doschek}, {Fletcher}, {Hara}, {Harra}, {Hillier},
  {Imada}, {Klimchuk}, {Mariska}, {Pereira}, {Reeves}, {Sakao}, {Sakurai},
  {Shimizu}, {Shimojo}, {Shiota}, {Solanki}, {Sterling}, {Su}, {Suematsu},
  {Tarbell}, {Tiwari}, {Toriumi}, {Ugarte-Urra}, {Warren}, {Watanabe}, \&
  {Young}}]{2019PASJ...71R...1H}
{Hinode Review Team}, {Al-Janabi}, K., {Antolin}, P., {et~al.} 2019, \pasj, 71,
  R1, \dodoi{10.1093/pasj/psz084}

\bibitem[{{Hou} {et~al.}(2017){Hou}, {Zhang}, {Li}, {Yang}, \&
  {Li}}]{2017ApJ...848L...9H}
{Hou}, Y., {Zhang}, J., {Li}, T., {Yang}, S., \& {Li}, X. 2017, \apjl, 848, L9,
  \dodoi{10.3847/2041-8213/aa8edd}

\bibitem[{{Hou} {et~al.}(2016{\natexlab{a}}){Hou}, {Li}, {Yang}, \&
  {Zhang}}]{2016A&A...589L...7H}
{Hou}, Y.~J., {Li}, T., {Yang}, S.~H., \& {Zhang}, J. 2016{\natexlab{a}}, \aap,
  589, L7, \dodoi{10.1051/0004-6361/201628216}

\bibitem[{{Hou} {et~al.}(2018){Hou}, {Huang}, {Xia}, {Li}, \&
  {Fu}}]{2018ApJ...855...65H}
{Hou}, Z., {Huang}, Z., {Xia}, L., {Li}, B., \& {Fu}, H. 2018, \apj, 855, 65,
  \dodoi{10.3847/1538-4357/aaab5a}

\bibitem[{{Hou} {et~al.}(2016{\natexlab{b}}){Hou}, {Huang}, {Xia}, {Li},
  {Madjarska}, \& {Fu}}]{2016AIPC.1720b0001H}
{Hou}, Z., {Huang}, Z., {Xia}, L., {et~al.} 2016{\natexlab{b}}, in American
  Institute of Physics Conference Series, Vol. 1720, Solar Wind 14, 020001,
  \dodoi{10.1063/1.4943802}

\bibitem[{{Huang} {et~al.}(2012){Huang}, {Madjarska}, {Doyle}, \&
  {Lamb}}]{2012A&A...548A..62H}
{Huang}, Z., {Madjarska}, M.~S., {Doyle}, J.~G., \& {Lamb}, D.~A. 2012, \aap,
  548, A62, \dodoi{10.1051/0004-6361/201220079}

\bibitem[{{Huang} {et~al.}(2015){Huang}, {Xia}, {Li}, \&
  {Madjarska}}]{2015ApJ...810...46H}
{Huang}, Z., {Xia}, L., {Li}, B., \& {Madjarska}, M.~S. 2015, \apj, 810, 46,
  \dodoi{10.1088/0004-637X/810/1/46}

\bibitem[{{Innes} {et~al.}(2015){Innes}, {Guo}, {Huang}, \&
  {Bhattacharjee}}]{2015ApJ...813...86I}
{Innes}, D.~E., {Guo}, L.~J., {Huang}, Y.~M., \& {Bhattacharjee}, A. 2015,
  \apj, 813, 86, \dodoi{10.1088/0004-637X/813/2/86}

\bibitem[{{Jiao} {et~al.}(2015){Jiao}, {Xia}, {Li}, {Huang}, {Li},
  {Chandrashekhar}, {Mou}, \& {Fu}}]{2015ApJ...809L..17J}
{Jiao}, F., {Xia}, L., {Li}, B., {et~al.} 2015, \apjl, 809, L17,
  \dodoi{10.1088/2041-8205/809/1/L17}

\bibitem[{{Kitagawa} \& {Yokoyama}(2015)}]{2015ApJ...805...97K}
{Kitagawa}, N., \& {Yokoyama}, T. 2015, \apj, 805, 97,
  \dodoi{10.1088/0004-637X/805/2/97}

\bibitem[{{Kosugi} {et~al.}(2007){Kosugi}, {Matsuzaki}, {Sakao}, {Shimizu},
  {Sone}, {Tachikawa}, {Hashimoto}, {Minesugi}, {Ohnishi}, {Yamada}, {Tsuneta},
  {Hara}, {Ichimoto}, {Suematsu}, {Shimojo}, {Watanabe}, {Shimada}, {Davis},
  {Hill}, {Owens}, {Title}, {Culhane}, {Harra}, {Doschek}, \&
  {Golub}}]{2007SoPh..243....3K}
{Kosugi}, T., {Matsuzaki}, K., {Sakao}, T., {et~al.} 2007, \solphys, 243, 3,
  \dodoi{10.1007/s11207-007-9014-6}

\bibitem[{{Lamb}(2008)}]{2008PhDT........20L}
{Lamb}, D.~A. 2008, PhD thesis, University of Colorado at Boulder

\bibitem[{{Lamb} {et~al.}(2008){Lamb}, {DeForest}, {Hagenaar}, {Parnell}, \&
  {Welsch}}]{2008ApJ...674..520L}
{Lamb}, D.~A., {DeForest}, C.~E., {Hagenaar}, H.~J., {Parnell}, C.~E., \&
  {Welsch}, B.~T. 2008, \apj, 674, 520, \dodoi{10.1086/524372}

\bibitem[{{Lamb} {et~al.}(2010){Lamb}, {DeForest}, {Hagenaar}, {Parnell}, \&
  {Welsch}}]{2010ApJ...720.1405L}
---. 2010, \apj, 720, 1405, \dodoi{10.1088/0004-637X/720/2/1405}

\bibitem[{{Lamb} {et~al.}(2014){Lamb}, {Howard}, \&
  {DeForest}}]{2014ApJ...788....7L}
{Lamb}, D.~A., {Howard}, T.~A., \& {DeForest}, C.~E. 2014, \apj, 788, 7,
  \dodoi{10.1088/0004-637X/788/1/7}

\bibitem[{{Lamb} {et~al.}(2013){Lamb}, {Howard}, {DeForest}, {Parnell}, \&
  {Welsch}}]{2013ApJ...774..127L}
{Lamb}, D.~A., {Howard}, T.~A., {DeForest}, C.~E., {Parnell}, C.~E., \&
  {Welsch}, B.~T. 2013, \apj, 774, 127, \dodoi{10.1088/0004-637X/774/2/127}

\bibitem[{{Leenaarts} {et~al.}(2013){Leenaarts}, {Pereira}, {Carlsson},
  {Uitenbroek}, \& {De Pontieu}}]{2013ApJ...772...90L}
{Leenaarts}, J., {Pereira}, T.~M.~D., {Carlsson}, M., {Uitenbroek}, H., \& {De
  Pontieu}, B. 2013, \apj, 772, 90, \dodoi{10.1088/0004-637X/772/2/90}

\bibitem[{{Lemen} {et~al.}(2012){Lemen}, {Title}, {Akin}, {Boerner}, {Chou},
  {Drake}, {Duncan}, {Edwards}, {Friedlaender}, {Heyman}, {Hurlburt}, {Katz},
  {Kushner}, {Levay}, {Lindgren}, {Mathur}, {McFeaters}, {Mitchell}, {Rehse},
  {Schrijver}, {Springer}, {Stern}, {Tarbell}, {Wuelser}, {Wolfson}, {Yanari},
  {Bookbinder}, {Cheimets}, {Caldwell}, {Deluca}, {Gates}, {Golub}, {Park},
  {Podgorski}, {Bush}, {Scherrer}, {Gummin}, {Smith}, {Auker}, {Jerram},
  {Pool}, {Soufli}, {Windt}, {Beardsley}, {Clapp}, {Lang}, \&
  {Waltham}}]{2012SoPh..275...17L}
{Lemen}, J.~R., {Title}, A.~M., {Akin}, D.~J., {et~al.} 2012, \solphys, 275,
  17, \dodoi{10.1007/s11207-011-9776-8}

\bibitem[{{Liu} \& {Su}(2014)}]{2014Ap&SS.351..417L}
{Liu}, S., \& {Su}, J.~T. 2014, \apss, 351, 417,
  \dodoi{10.1007/s10509-014-1853-7}

\bibitem[{{Mandrini} {et~al.}(2014){Mandrini}, {Nuevo}, {V{\'a}squez},
  {D{\'e}moulin}, {van Driel-Gesztelyi}, {Baker}, {Culhane}, {Cristiani}, \&
  {Pick}}]{2014SoPh..289.4151M}
{Mandrini}, C.~H., {Nuevo}, F.~A., {V{\'a}squez}, A.~M., {et~al.} 2014,
  \solphys, 289, 4151, \dodoi{10.1007/s11207-014-0582-y}

\bibitem[{{Marsch} {et~al.}(2008){Marsch}, {Tian}, {Sun}, {Curdt}, \&
  {Wiegelmann}}]{2008ApJ...685.1262M}
{Marsch}, E., {Tian}, H., {Sun}, J., {Curdt}, W., \& {Wiegelmann}, T. 2008,
  \apj, 685, 1262, \dodoi{10.1086/591038}

\bibitem[{{Marsch} {et~al.}(2004){Marsch}, {Wiegelmann}, \&
  {Xia}}]{2004A&A...428..629M}
{Marsch}, E., {Wiegelmann}, T., \& {Xia}, L.~D. 2004, \aap, 428, 629,
  \dodoi{10.1051/0004-6361:20041060}

\bibitem[{{M{\"u}ller} {et~al.}(2020){M{\"u}ller}, {St. Cyr}, {Zouganelis},
  {Gilbert}, {Marsden}, {Nieves-Chinchilla}, {Antonucci}, {Auch{\`e}re},
  {Berghmans}, {Horbury}, {Howard}, {Krucker}, {Maksimovic}, {Owen}, {Rochus},
  {Rodriguez-Pacheco}, {Romoli}, {Solanki}, {Bruno}, {Carlsson}, {Fludra},
  {Harra}, {Hassler}, {Livi}, {Louarn}, {Peter}, {Sch{\"u}hle}, {Teriaca}, {del
  Toro Iniesta}, {Wimmer-Schweingruber}, {Marsch}, {Velli}, {De Groof},
  {Walsh}, \& {Williams}}]{2020A&A...642A...1M}
{M{\"u}ller}, D., {St. Cyr}, O.~C., {Zouganelis}, I., {et~al.} 2020, \aap, 642,
  A1, \dodoi{10.1051/0004-6361/202038467}

\bibitem[{{Nishizuka} \& {Hara}(2011)}]{2011ApJ...737L..43N}
{Nishizuka}, N., \& {Hara}, H. 2011, \apjl, 737, L43,
  \dodoi{10.1088/2041-8205/737/2/L43}

\bibitem[{{Pant} {et~al.}(2015){Pant}, {Dolla}, {Mazumder}, {Banerjee},
  {Krishna Prasad}, \& {Panditi}}]{2015ApJ...807...71P}
{Pant}, V., {Dolla}, L., {Mazumder}, R., {et~al.} 2015, \apj, 807, 71,
  \dodoi{10.1088/0004-637X/807/1/71}

\bibitem[{{Parnell} {et~al.}(2009){Parnell}, {DeForest}, {Hagenaar},
  {Johnston}, {Lamb}, \& {Welsch}}]{2009ApJ...698...75P}
{Parnell}, C.~E., {DeForest}, C.~E., {Hagenaar}, H.~J., {et~al.} 2009, \apj,
  698, 75, \dodoi{10.1088/0004-637X/698/1/75}

\bibitem[{{Pereira} {et~al.}(2013){Pereira}, {Leenaarts}, {De Pontieu},
  {Carlsson}, \& {Uitenbroek}}]{2013ApJ...778..143P}
{Pereira}, T.~M.~D., {Leenaarts}, J., {De Pontieu}, B., {Carlsson}, M., \&
  {Uitenbroek}, H. 2013, \apj, 778, 143, \dodoi{10.1088/0004-637X/778/2/143}

\bibitem[{{Pesnell} {et~al.}(2012){Pesnell}, {Thompson}, \&
  {Chamberlin}}]{2012SoPh..275....3P}
{Pesnell}, W.~D., {Thompson}, B.~J., \& {Chamberlin}, P.~C. 2012, \solphys,
  275, 3, \dodoi{10.1007/s11207-011-9841-3}

\bibitem[{{Polito} {et~al.}(2020){Polito}, {De Pontieu}, {Testa}, {Brooks}, \&
  {Hansteen}}]{2020ApJ...903...68P}
{Polito}, V., {De Pontieu}, B., {Testa}, P., {Brooks}, D.~H., \& {Hansteen}, V.
  2020, \apj, 903, 68, \dodoi{10.3847/1538-4357/abba1d}

\bibitem[{{Ruan} {et~al.}(2016){Ruan}, {He}, {Zhang}, {Vocks}, {Marsch}, {Tu},
  {Peter}, \& {Wang}}]{2016ApJ...825...58R}
{Ruan}, W., {He}, J., {Zhang}, L., {et~al.} 2016, \apj, 825, 58,
  \dodoi{10.3847/0004-637X/825/1/58}

\bibitem[{{Sakao} {et~al.}(2007){Sakao}, {Kano}, {Narukage}, {Kotoku}, {Bando},
  {DeLuca}, {Lundquist}, {Tsuneta}, {Harra}, {Katsukawa}, {Kubo}, {Hara},
  {Matsuzaki}, {Shimojo}, {Bookbinder}, {Golub}, {Korreck}, {Su}, {Shibasaki},
  {Shimizu}, \& {Nakatani}}]{2007Sci...318.1585S}
{Sakao}, T., {Kano}, R., {Narukage}, N., {et~al.} 2007, Science, 318, 1585,
  \dodoi{10.1126/science.1147292}

\bibitem[{{Scherrer} {et~al.}(2012){Scherrer}, {Schou}, {Bush}, {Kosovichev},
  {Bogart}, {Hoeksema}, {Liu}, {Duvall}, {Zhao}, {Title}, {Schrijver},
  {Tarbell}, \& {Tomczyk}}]{2012SoPh..275..207S}
{Scherrer}, P.~H., {Schou}, J., {Bush}, R.~I., {et~al.} 2012, \solphys, 275,
  207, \dodoi{10.1007/s11207-011-9834-2}

\bibitem[{{Slemzin} {et~al.}(2013){Slemzin}, {Harra}, {Urnov}, {Kuzin},
  {Goryaev}, \& {Berghmans}}]{2013SoPh..286..157S}
{Slemzin}, V., {Harra}, L., {Urnov}, A., {et~al.} 2013, \solphys, 286, 157,
  \dodoi{10.1007/s11207-012-0004-y}

\bibitem[{{Spice Consortium} {et~al.}(2020){Spice Consortium}, {Anderson},
  {Appourchaux}, {Auch{\`e}re}, {Aznar Cuadrado}, {Barbay}, {Baudin},
  {Beardsley}, {Bocchialini}, {Borgo}, {Bruzzi}, {Buchlin}, {Burton},
  {B{\"u}chel}, {Caldwell}, {Caminade}, {Carlsson}, {Curdt}, {Davenne},
  {Davila}, {Deforest}, {Del Zanna}, {Drummond}, {Dubau}, {Dumesnil}, {Dunn},
  {Eccleston}, {Fludra}, {Fredvik}, {Gabriel}, {Giunta}, {Gottwald}, {Griffin},
  {Grundy}, {Guest}, {Gyo}, {Haberreiter}, {Hansteen}, {Harrison}, {Hassler},
  {Haugan}, {Howe}, {Janvier}, {Klein}, {Koller}, {Kucera}, {Kouliche},
  {Marsch}, {Marshall}, {Marshall}, {Matthews}, {McQuirk}, {Meining},
  {Mercier}, {Morris}, {Morse}, {Munro}, {Parenti}, {Pastor-Santos}, {Peter},
  {Pfiffner}, {Phelan}, {Philippon}, {Richards}, {Rogers}, {Sawyer},
  {Schlatter}, {Schmutz}, {Sch{\"u}hle}, {Shaughnessy}, {Sidher}, {Solanki},
  {Speight}, {Spescha}, {Szwec}, {Tamiatto}, {Teriaca}, {Thompson}, {Tosh},
  {Tustain}, {Vial}, {Walls}, {Waltham}, {Wimmer-Schweingruber}, {Woodward},
  {Young}, {de Groof}, {Pacros}, {Williams}, \&
  {M{\"u}ller}}]{2020A&A...642A..14S}
{Spice Consortium}, {Anderson}, M., {Appourchaux}, T., {et~al.} 2020, \aap,
  642, A14, \dodoi{10.1051/0004-6361/201935574}

\bibitem[{{Srivastava} {et~al.}(2014){Srivastava}, {Konkol}, {Murawski},
  {Dwivedi}, \& {Mohan}}]{2014SoPh..289.4501S}
{Srivastava}, A.~K., {Konkol}, P., {Murawski}, K., {Dwivedi}, B.~N., \&
  {Mohan}, A. 2014, \solphys, 289, 4501, \dodoi{10.1007/s11207-014-0584-9}

\bibitem[{{Testa} {et~al.}(2020){Testa}, {Polito}, \& {De
  Pontieu}}]{2020ApJ...889..124T}
{Testa}, P., {Polito}, V., \& {De Pontieu}, B. 2020, \apj, 889, 124,
  \dodoi{10.3847/1538-4357/ab63cf}

\bibitem[{{Tian} {et~al.}(2021){Tian}, {Harra}, {Baker}, {Brooks}, \&
  {Xia}}]{2021SoPh..296...47T}
{Tian}, H., {Harra}, L., {Baker}, D., {Brooks}, D.~H., \& {Xia}, L. 2021,
  \solphys, 296, 47, \dodoi{10.1007/s11207-021-01792-7}

\bibitem[{{Tian} {et~al.}(2011){Tian}, {McIntosh}, \& {De
  Pontieu}}]{2011ApJ...727L..37T}
{Tian}, H., {McIntosh}, S.~W., \& {De Pontieu}, B. 2011, \apjl, 727, L37,
  \dodoi{10.1088/2041-8205/727/2/L37}

\bibitem[{{Tian} {et~al.}(2012){Tian}, {McIntosh}, {Wang}, {Ofman}, {De
  Pontieu}, {Innes}, \& {Peter}}]{2012ApJ...759..144T}
{Tian}, H., {McIntosh}, S.~W., {Wang}, T., {et~al.} 2012, \apj, 759, 144,
  \dodoi{10.1088/0004-637X/759/2/144}

\bibitem[{{Tsuneta} {et~al.}(2008){Tsuneta}, {Ichimoto}, {Katsukawa}, {Nagata},
  {Otsubo}, {Shimizu}, {Suematsu}, {Nakagiri}, {Noguchi}, {Tarbell}, {Title},
  {Shine}, {Rosenberg}, {Hoffmann}, {Jurcevich}, {Kushner}, {Levay}, {Lites},
  {Elmore}, {Matsushita}, {Kawaguchi}, {Saito}, {Mikami}, {Hill}, \&
  {Owens}}]{2008SoPh..249..167T}
{Tsuneta}, S., {Ichimoto}, K., {Katsukawa}, Y., {et~al.} 2008, \solphys, 249,
  167, \dodoi{10.1007/s11207-008-9174-z}

\bibitem[{{Tu} {et~al.}(2005){Tu}, {Zhou}, {Marsch}, {Xia}, {Zhao}, {Wang}, \&
  {Wilhelm}}]{2005Sci...308..519T}
{Tu}, C.-Y., {Zhou}, C., {Marsch}, E., {et~al.} 2005, Science, 308, 519,
  \dodoi{10.1126/science.1109447}

\bibitem[{{Ugarte-Urra} \& {Warren}(2011)}]{2011ApJ...730...37U}
{Ugarte-Urra}, I., \& {Warren}, H.~P. 2011, \apj, 730, 37,
  \dodoi{10.1088/0004-637X/730/1/37}

\bibitem[{{van Driel-Gesztelyi} {et~al.}(2012){van Driel-Gesztelyi}, {Culhane},
  {Baker}, {D{\'e}moulin}, {Mandrini}, {DeRosa}, {Rouillard}, {Opitz},
  {Stenborg}, {Vourlidas}, \& {Brooks}}]{2012SoPh..281..237V}
{van Driel-Gesztelyi}, L., {Culhane}, J.~L., {Baker}, D., {et~al.} 2012,
  \solphys, 281, 237, \dodoi{10.1007/s11207-012-0076-8}

\bibitem[{{Vanninathan} {et~al.}(2015){Vanninathan}, {Madjarska}, {Galsgaard},
  {Huang}, \& {Doyle}}]{2015A&A...584A..38V}
{Vanninathan}, K., {Madjarska}, M.~S., {Galsgaard}, K., {Huang}, Z., \&
  {Doyle}, J.~G. 2015, \aap, 584, A38, \dodoi{10.1051/0004-6361/201526340}

\bibitem[{{Warren} {et~al.}(2011){Warren}, {Ugarte-Urra}, {Young}, \&
  {Stenborg}}]{2011ApJ...727...58W}
{Warren}, H.~P., {Ugarte-Urra}, I., {Young}, P.~R., \& {Stenborg}, G. 2011,
  \apj, 727, 58, \dodoi{10.1088/0004-637X/727/1/58}

\bibitem[{{Wiegelmann} {et~al.}(2014){Wiegelmann}, {Thalmann}, \&
  {Solanki}}]{2014A&ARv..22...78W}
{Wiegelmann}, T., {Thalmann}, J.~K., \& {Solanki}, S.~K. 2014, \aapr, 22, 78,
  \dodoi{10.1007/s00159-014-0078-7}

\bibitem[{{W{\"u}lser} {et~al.}(2018){W{\"u}lser}, {Jaeggli}, {De Pontieu},
  {Tarbell}, {Boerner}, {Freeland}, {Liu}, {Timmons}, {Brannon}, {Kankelborg},
  {Madsen}, {McKillop}, {Prchlik}, {Saar}, {Schanche}, {Testa}, {Bryans}, \&
  {Wiesmann}}]{2018SoPh..293..149W}
{W{\"u}lser}, J.~P., {Jaeggli}, S., {De Pontieu}, B., {et~al.} 2018, \solphys,
  293, 149, \dodoi{10.1007/s11207-018-1364-8}

\bibitem[{{Yang} {et~al.}(2015){Yang}, {Zhang}, {Jiang}, \&
  {Xiang}}]{2015ApJ...804L..27Y}
{Yang}, S., {Zhang}, J., {Jiang}, F., \& {Xiang}, Y. 2015, \apjl, 804, L27,
  \dodoi{10.1088/2041-8205/804/2/L27}

\bibitem[{{Young} {et~al.}(2012){Young}, {O'Dwyer}, \&
  {Mason}}]{2012ApJ...744...14Y}
{Young}, P.~R., {O'Dwyer}, B., \& {Mason}, H.~E. 2012, \apj, 744, 14,
  \dodoi{10.1088/0004-637X/744/1/14}

\bibitem[{{Zhang} {et~al.}(2015){Zhang}, {He}, {Yan}, {Tu}, {Wang}, \&
  {Wang}}]{2015ScChD..58..830Z}
{Zhang}, J., {He}, J., {Yan}, L., {et~al.} 2015, Science China Earth Sciences,
  58, 830, \dodoi{10.1007/s11430-014-4999-9}

\bibitem[{{Zhang} {et~al.}(2017){Zhang}, {Tian}, {He}, \&
  {Wang}}]{2017ApJ...838....2Z}
{Zhang}, J., {Tian}, H., {He}, J., \& {Wang}, L. 2017, \apj, 838, 2,
  \dodoi{10.3847/1538-4357/aa63e8}

\end{thebibliography}
\bibliographystyle{aasjournal}

\end{document}